\documentclass[preprint,trackchanges,twocolumn]{aastex7}
\usepackage{tabularx, makecell}
\usepackage{graphicx}	
\usepackage{amsmath}	
\usepackage[T1]{fontenc}
\usepackage{newtxtext,newtxmath}

\usepackage{textcomp}
\usepackage{makecell}
\usepackage{array}
\usepackage{graphicx}
\usepackage[utf8]{inputenc} 
\usepackage{textcomp}
\usepackage{xcolor}

\begin{document}

\title{Constraining Spin and Inclination Angle of XTE J2012+381 using AstroSat and NICER}

\author[orcid=0009-0003-1029-5201,sname='Sharma']{Vaibhav Sharma}
\affiliation{Department of Physics, Indian Institute of Technology Kanpur, Kanpur, Uttar Pradesh - 208016, India}
\email[show]{svbhv@iitk.ac.in}  

\author[gname=Ranjeev, sname='Misra']{Ranjeev Misra}
\affiliation{Inter-University Center for Astronomy and Astrophysics, Ganeshkhind, Pune, Maharashtra - 411007, India}
\email{rmisra@iucaa.in}
\author[gname=Shivani,sname=Chaudhary]{Shivani Chaudhary}
\affiliation{Department of Physics, Banasthali Vidyapith, Rajasthan - 304022, India}
\email[]{shivani.chaudhary.research@gmail.com}
\author[gname=J S,sname=Yadav]{J S Yadav}
\affiliation{Space, Planetary and Astronomical Sciences and Engineering, IIT Kanpur, Kanpur Nagar, Uttar Pradesh - 208016, India}
\affiliation{Department of Astronomy and Astrophysics, Tata Institute of Fundamental Research, Mumbai, Maharashtra - 400005, India}
\email{jsyadav@iitk.ac.in}

\author{Pankaj Jain}
\affiliation{Space, Planetary and Astronomical Sciences and Engineering, IIT Kanpur, Kanpur Nagar, Uttar Pradesh - 208016, India}
\email{pkjain@iitk.ac.in}

\begin{abstract}

We present a spectral analysis of a black hole X-ray binary XTE J2012+381 during its 2022 outburst, using data from NICER and AstroSat. Combining data from NICER, LAXPC20, and SXT, we extract energy spectra covering the 0.7–10.0 keV range. We model the energy spectra using a series of physical models and find that a reflection–Comptonization model provides the best fit. Given the uncertainties in the black hole mass and source distance, we investigate the stability of the inferred spectral parameters by systematically varying the black hole mass (7.26, 11, and 16.5 M$_\odot$), source distance (3.3, 5.4, and 7.5 kpc), and spectral hardening factor (1.5, 1.7, and 1.9). We find that, across most combinations of these parameters, the spin solutions consistently lie in the high-spin regime, spanning values between $\sim$0.67 and $\sim$0.998, with only a limited subset of configurations favoring lower spins. In contrast, the disk inclination angle remains well constrained over the majority of the explored parameter space, typically ranging between $\sim$50° and $\sim$65°. Only a few parameter combinations yield higher inclination values.

\end{abstract}



\section{INTRODUCTION}
The observed spin distribution of stellar-mass black holes (BHs) in X-ray binaries (XRBs) significantly differs from that of black holes in merging binary black hole (BBH) systems observed through gravitational waves (GWs). While BHs in XRBs tend to exhibit high spins, those in BBHs preferentially show low spins \citep{Fishbach...2022,Draghis...2023b}. Spin values in XRBs are typically reported with only statistical uncertainties, as systematic uncertainties remain poorly understood. Hence, this is necessary to comprehend whether it is a physical difference or it is caused by our selection effects.

To address this, it is crucial to expand the sample of spin measurements obtained through reliable and independent methods. Two well-established X-ray techniques are commonly used to estimate black hole spin. Continuum-fitting \citep{Gou...2009, Feng...2023, Zhang...1997}, models the thermal emission from the accretion disk, requires independent knowledge of the black hole’s mass, distance, and disk inclination angle. Relativistic reflection modeling (also called the iron line method) \citep{Fabian...1989, Tanaka...1995, Fabian...2000, Brenneman...2006, Miller...2007, Miller...2010, Draghis...2020}, analyzes spectral features produced when hard X-rays from the corona reflect off the accretion disk. Thermal photons emitted from the inner region of the accretion disk get inverse comptonized by the highly energetic electrons present in the corona. Some fraction of these inversely comptonized photons reprocess through the accretion disk and produce an iron line and reflection hump in the energy spectrum. This approach does not require external measurements of mass or distance.

XTE~J2012+381 is classified as a black hole X-ray binary candidate based on its spectral and timing properties observed during the 1998 outburst. The source was initially identified with the \textit{Rossi X-ray Timing Explorer} (RXTE) All-Sky Monitor \citep{Remillard...1998}. Spectral modeling of follow-up \textit{ASCA} observations revealed features characteristic of accretion onto a black hole \citep{White...1998}, while analysis of RXTE data indicated the presence of a broadened Fe K$\alpha$ emission line near $\sim 6.4$~keV, suggestive of reflection from the inner accretion disk \citep{Vasiliev...2000}. Constraints on the compact object mass were later derived as a function of source distance, giving lower limits of $M_{\mathrm{BH}} \geq 22\, d_{10}\, M_\odot$ for a maximally spinning black hole and $M_{\mathrm{BH}} \geq 3.7\, d_{10}\, M_\odot$ for a non-rotating black hole, where $d_{10}$ denotes the distance in units of 10~kpc \citep{Campana...2002}. Independent parallax measurements from \textit{Gaia} subsequently estimated the source distance to be $5.4^{+2.1}_{-2.1}$~kpc \citep{Gaia...2016}.

In 2022, the source underwent another outburst. MAXI \citep{MAXI...ATel} and Swift \citep{SWIFT...ATel} reported this new outburst, and follow-up observations were triggered with NICER \citep{NICER...ATel} and AstroSat. \cite{Draghis...2023a} analyzed the 2022 outburst of the XTE J2012+381 using 105 NICER and 2 NuSTAR observations. They found clear evidence of relativistic disk reflection in the NuSTAR data, allowing them to measure a very high black hole spin of a = $0.988^{+0.008}_{-0.030}$ and a disk inclination of $68^{+6}_{-11}$ degrees with one sigma statistical errors. \cite{Raj...2024} presented a detailed X-ray spectral analysis of the XTE J2012+381 during its 2022 outburst using data from Swift/XRT and NuSTAR. This work measured the black hole's spin and mass by applying both the relxillCp and relxilllpCp relativistic reflection models. For a high disk density ($10^{20}$ cm$^{-3}$), this work found the spin of $0.883^{+0.033}_{-0.061}$ and an inclination angle of $46.2^{+3.7}_{-2.0}$ degrees using the relxillCp model and the spin of $0.892^{+0.020}_{-0.044}$ and an inclination angle of $43.1^{+1.4}_{-1.2}$ degrees using relxillpCp.  

Simultaneous observations of AstroSat and NICER provide an opportunity to understand the effects of spectral parameters such as spectral hardening factor, black hole mass, and the source distance on the estimation of spin parameter and inclination angle.

In Section~\ref{sec: obs and data red}, we describe the instruments used, the observation log, and the data reduction procedure. Section~\ref{sec:Analysis and Results} presents the details of the spectral analysis. The interpretation of our findings and conclusions is discussed in Section~\ref{sec:Discussion}.

\section{Observations and Data reduction} \label{sec: obs and data red}
NICER monitored the source evolution during the 2022 outburst through a total of 143 observations. On MJD 60031 (March 28, 2023), AstroSat and NICER observed the source, enabling simultaneous coverage. Though there are 143 NICER observations, the analysis in this work only uses the one simultaneous with the AstroSat observation. A detailed observation log is provided in Table~\ref{tab:obslog}.

AstroSat, India’s first dedicated astronomy satellite, enables simultaneous multi-wavelength observations \citep{Singh_2014} from X-rays to the ultraviolet using five onboard payloads. The primary instruments used in this study are the Large Area X-ray Proportional Counters (LAXPC) \citep{Yadav...2016, Agarwal...2017, Antia...2017} and the Soft X-ray Telescope (SXT) \citep{Singh...2016, Singh...2017}. The LAXPC comprises three identical units (LAXPC10, LAXPC20, and LAXPC30), each covering an effective area of approximately 6000 cm$^2$ at 15~keV, a timing resolution of 10~$\mu$s, and a broad energy coverage from 3.0 to 80.0~keV. The SXT provides an effective area of 90~cm$^2$ at 1.5~keV, with timing resolution up to 2.37~s and an energy range spanning 0.3–8.0~keV.

For this work, we analyze data from AstroSat’s LAXPC20 and SXT instruments. The LAXPC20 level-1 data are downloaded from the \href{https://astrobrowse.issdc.gov.in/astro_archive/archive/Home.jsp}{AstroSat data archive\footnote{\url{https://astrobrowse.issdc.gov.in/astro_archive/archive/Home.jsp}}}. We process the data using the \href{https://www.tifr.res.in/~astrosat_laxpc/LaxpcSoft.html}{LAXPC software\footnote{\url{https://www.tifr.res.in/~astrosat_laxpc/LaxpcSoft.html}}} to generate the \texttt{level2.event.fits} file, which is used further to extract science products. Due to gain instability and known response issues in LAXPC10 and LAXPC30 \citep{Antia...2017, Antia...2021}, we exclude data from these units.

From the LAXPC20 \texttt{level2.event.fits} file, we extract light curves and energy spectra using data from Layer 1 only. The power density spectra (PDS) is extracted using all layers in the 4.0–10.0~keV energy band.

The SXT level-2 data are also retrieved from the \href{https://astrobrowse.issdc.gov.in/astro_archive/archive/Home.jsp}{AstroSat data archive}. We merge data from all available orbits using the \href{https://github.com/gulabd/SXTMerger.jl}{SXT event merger\footnote{\url{https://github.com/gulabd/SXTMerger.jl}}} tool, developed in Julia. We use the merged clean event file to extract light curves and spectra in the 0.7–4.0~keV range. The source region is selected using the DS9 tool, and background subtraction is done using the background file \texttt{SkyBkg\_comb\_EL3p5\_Cl\_Rd16p0\_v01.pha} provided by the \href{https://www.tifr.res.in/~astrosat_sxt/dataanalysis.html}{AstroSat SXT science support cell\footnote{\url{https://www.tifr.res.in/~astrosat_sxt/dataanalysis.html}}}. For response calibration, we use the file \texttt{sxt\_pc\_mat\_g0to12.rmf}, and apply vignetting correction to the ARF using the \href{https://www.tifr.res.in/~astrosat_sxt/dataanalysis.html}{SXTARFModule\_v02 tool\footnote{\url{https://www.tifr.res.in/~astrosat_sxt/dataanalysis.html}}}.

All spectral files, the energy spectrum, the response matrix file, and the background spectrum are grouped using the \texttt{ftgrppha} tool with an optimal binning scheme as recommended by \citet{Kaastra...2016}.

NICER is an X-ray telescope mounted on the International Space Station (ISS), optimized for soft X-ray observations. It operates in the 0.2–12.0~keV energy range and offers an effective area of approximately 1900~cm$^2$ at 1.5~keV. NICER provides excellent timing resolution of 100 nanoseconds and a spectral resolution of $\sim$85~eV at 1~keV.

We process the NICER observation using the standard pipeline. The \texttt{nicerl2} task is used to produce the \texttt{clean.evt} file, which serves as the input for spectrum, response, and background file generation using the \texttt{nicerl3-spect} task. Light curves are extracted using the \texttt{nicerl3-lc} task. For background estimation, we create the \texttt{3C50} background using \texttt{nibackgen3C50} command. Due to optical loading dominance at low energies \citep{Draghis...2023a} in the NICER observations, we use NICER data in the 1.0-8.0 keV energy range in our spectral analysis.

\begin{table}[ht]
\centering

\begin{tabularx}{\columnwidth}{l c l@{}}
    \hline\hline
    Observation ID & 
    \makecell[c]{\hspace{-1.5em}Obs. Date\\ \hspace{-1.5em}(MJD)} & 
    \makecell[l]{Exposure \\ Time (ks)} \\
    \hline
    T05\_070T01\_9000005544\textsuperscript{\textbf{*}} & \hspace{1.0em}60030.16 & \hspace{3.0em}$\sim$23.53 \\
    6203600121\textsuperscript{\textbf{$\dagger$}} & \hspace{1.0em}60031.01 & \hspace{3.0em}$\sim$1.63 \\
    \hline
\end{tabularx}
\caption{Observation log of AstroSat\textsuperscript{\textbf{*}} and NICER\textsuperscript{\textbf{$\dagger$}}  used in this work.}

\label{tab:obslog}
\end{table}

\section{Spectral Analysis and Results} \label{sec:Analysis and Results}

To understand the origin of the X-ray emission from the source, we start with a phenomenological approach. We assume that the observed X-ray spectrum comprises two primary components: thermal photons are directly emitted from the accretion disk, and non-thermal photons are a result of inverse Compton scattering of these disk photons by a hot corona. Accordingly, we start with an initial model in \textsc{XSPEC} using the form: \texttt{constant} $\times$ \texttt{tbabs} $\times$ (\texttt{thcomp} $\times$ \texttt{diskbb}). Here, \texttt{diskbb} \citep{Mitsuda...1984} represents the multi-temperature blackbody emission from the accretion disk, \texttt{thcomp} \citep{Zdziarski2020} models thermal Comptonization by the corona, and \texttt{tbabs} \citep{Wilms2000} accounts for interstellar absorption. A \texttt{constant} factor is included to allow for cross-normalization among instruments. We use simultaneous data from SXT (0.7–4.0~keV), LAXPC (4.0–10.0~keV), and NICER (1.0–8.0~keV).

During spectral fitting, we notice strong residuals near 6.4~keV, indicative of an iron K$\alpha$ emission line. To account for this, we include a \texttt{Gaussian} component to model the iron line. Hence, our model becomes \texttt{constant} $\times$ \texttt{tbabs} $\times$ (\texttt{gaussian} + \texttt{thcomp} $\times$ \texttt{diskbb}). Additionally, we use the \texttt{gain fit} command in \textsc{XSPEC} to correct for possible gain shifts, particularly in the SXT data. Assuming a 2\% systematic error, this model (referred to as Model~1S) results in a statistically acceptable fit with a chi-square of 99.40 for 154 degrees of freedom. 

To move toward a more physically motivated representation, we replace the empirical components with self-consistent models. Specifically, we substitute the \texttt{Gaussian} component with the \texttt{relxillCp} model \citep{Garcia2014}, which self-consistently describes both the thermal Comptonization and the relativistic reflection from the ionized accretion disk. Additionally, we replace the \texttt{diskbb} component with \texttt{kerrbb} \citep{Li...2005}, a fully relativistic disk model that incorporates the effects of black hole spin and general relativity. The revised model takes the form of \texttt{constant} $\times$ \texttt{tbabs} $\times$ (\texttt{relxillCp} + \texttt{thcomp} $\times$ \texttt{kerrbb}), hereafter referred to as Model~2S.

Based on the estimates reported by \cite{Campana...2002} and \cite{Gaia...2016}, the black hole mass is fixed at 11 M$_\odot$ and the source distance is fixed at 5.4 kpc in this model. In addition, the electron temperature is fixed at 150 keV, and the normalization of \texttt{kerrbb} is set to unity. We allow the hydrogen column density (nH), inclination angle ($i$), spin parameter (a), photon index ($\Gamma$), ionization parameter ($\log\xi$), iron abundance (A$_{Fe}$), covering fraction ($f_{sc}$), and mass accretion rate ($\dot{M}$) to vary during the fit. To maintain physical consistency, we tie the inclination and spin parameters between \texttt{kerrbb} and \texttt{relxillCp}, and the photon index and electron temperature between \texttt{thcomp} and \texttt{relxillCp}. This model also provides a statistically good and better fit, yielding a chi-square of 86.98 for 152 degrees of freedom. It provides meaningful physical constraints, with the black hole spin estimated to be $a = 0.87^{+0.03}_{-0.02}$ and the inclination angle as $58.06^{+3.65}_{-2.36}$ degrees.

\begin{figure}
\centering
	\includegraphics[width=1.15\columnwidth, height=6.2cm]{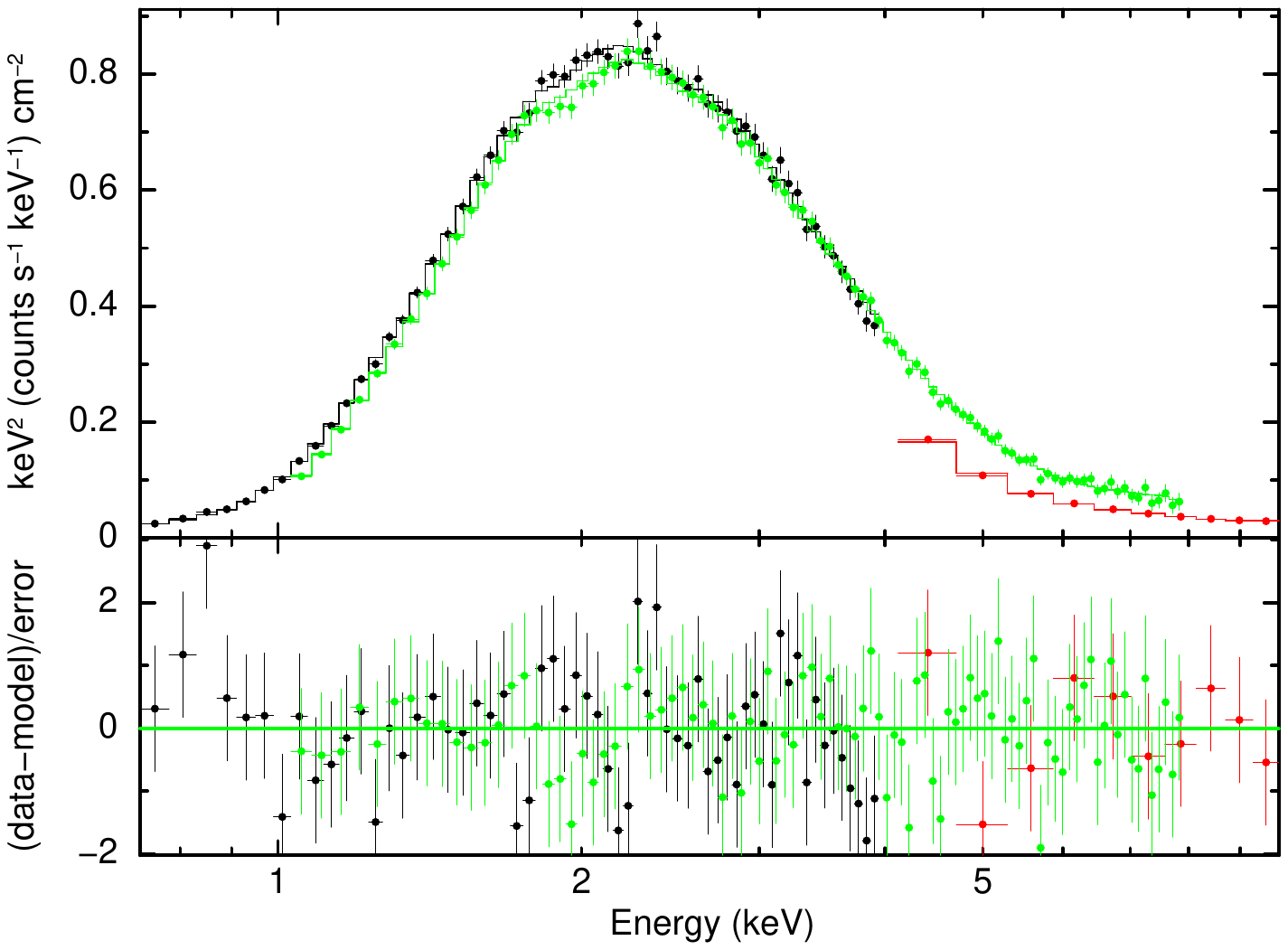}
        \caption{Joint spectral fitting of the energy spectra from AstroSat/SXT (black), AstroSat/LAXPC (red), and NICER (green) using the final model, \texttt{constant} $\times$ \texttt{tbabs} $\times$ (\texttt{relconv} $\times$ \texttt{xilconv} $\times$ \texttt{thcomp} $\times$ \texttt{kerrbb} + \texttt{thcomp} $\times$ \texttt{kerrbb}). The upper panel shows the energy spectra along with the best-fit model, while the lower panel displays the $\chi$.}
        \label{fig:EnergySpectrum}
\end{figure}

Both the continuum-fitting and relativistic reflection methods rely on the assumption that the accretion disk extends down to the innermost stable circular orbit (ISCO). In the present work, the source spectrum is strongly disk-dominated (Figure \ref{fig:EnergySpectrum} $\&$ Table \ref{Tab: All Spectral Parameters}), characterized by a prominent thermal component and a relatively weak Comptonized tail, indicating that the source was in the soft spectral state \citep{Remillard_2006} during this observation. Previous observational studies have shown that, in the soft state, the accretion disk in black hole X-ray binaries is expected to extend close to the ISCO (e.g., \citealt{Done_2007}).

Although Model 2S provided a good statistical fit, it has some inherent limitations. In particular, the \texttt{relxillCp model}, while self-consistent in handling reflection and Comptonization, assumes that the input seed photon temperature for the Comptonizing medium is at 10 eV, which is inconsistent with the significantly higher disc temperature. Moreover, the model combination does not provide an estimation of the reflection fraction. To overcome these issues and better reflect the physical geometry of the system, we construct a final model (Model~3S) that treats each component of the emission as a disk, corona, reflection, and relativistic distortion in a physically motivated way.

The model is structured as:
\begin{center}
\texttt{constant} $\times$ \texttt{tbabs} $\times$ (\texttt{constant} $\times$ \texttt{relconv} $\times$ \texttt{xilconv} $\times$ \texttt{thcomp} $\times$ \texttt{kerrbb} + \texttt{thcomp} $\times$ \texttt{kerrbb})
\end{center}

\begin{table}
\centering
\caption{Best-fit spectral parameters after fitting with XSPEC model \texttt{constant} $\times$ \texttt{tbabs} $\times$ (\texttt{relconv} $\times$ \texttt{xilconv} $\times$ \texttt{thcomp} $\times$ \texttt{kerrbb} + \texttt{thcomp} $\times$ \texttt{kerrbb}). Frozen parameters are marked with a dagger (†). Errors are provided with a 90$\%$ confidence level.}
\label{Tab: All Spectral Parameters}
\begin{tabular}{lccll}
\hline\hline
Par & Comp. & Parameter & Value & Note \\
\hline
1 & constant & factor & $1.000^{\dagger}$ & SXT \\
2 & TBabs    & nH ($10^{22}$ cm$^{-2}$) & $1.74^{+0.02}_{-0.02}$ & \\
3 & constant & factor & $1.63 \times10^{-3}$ & = p34$\times$(1-e$^{-p32}$) \\
4 & relconv  & Index1 & $3.0^{\dagger}$ & \\
5 & relconv  & Index2 & $3.0^{\dagger}$ & \\
6 & relconv  & $R_{\rm br}$ & $399^{\dagger}$ & \\
7 & relconv  & $a$ & $0.91^{+0.03}_{-0.05}$ & \\
8 & relconv  & Incl (deg) & $57.40^{+3.13}_{-3.47}$ & \\
9 & relconv  & $R_{\rm in}$ & $-1.0^{\dagger}$ & \\
10 & relconv  & $R_{\rm out}$ & $400^{\dagger}$ & \\
11 & relconv  & limb & $0.0^{\dagger}$ & \\
12 & xilconv  & rel\_refl & $<-13.67$ & \\
13 & xilconv  & redshift & $0.0^{\dagger}$ & \\
14 & xilconv  & Fe\_abund & $<0.90$ & \\
15 & xilconv  & cosIncl & $0.520^{\dagger}$ & = cosd(p8) \\
16 & xilconv  & log$\xi$ & $2.70^{+0.18}_{-0.54}$ & \\
17 & xilconv  & cutoff (keV) & $300^{\dagger}$ & \\
18 & thcomp   & $\tau$ & $0.40$ & = p32 \\
19 & thcomp   & $kT_e$ (keV) & $150^{\dagger}$ & = p33\\
20 & thcomp   & cov\_frac & $1.0^{\dagger}$ &  \\
21 & thcomp   & $z$ & $0.0^{\dagger}$ & \\
22 & kerrbb   & $\eta$ & $0.0$ & = p36\\
23 & kerrbb   & $a$ & $0.91$  & = p37 \\
24 & kerrbb   & $i$ (deg) & $57.40$ & = p38\\
25 & kerrbb   & $M_{\rm BH}$ ($M_\odot$) & $11.0$ & = p39 \\
26 & kerrbb   & $\dot{M}$ & $0.141$ & = p40 \\
27 & kerrbb   & $D_{\rm BH}$ (kpc) & $5.4$ & = p41\\
28 & kerrbb   & $hd$ & $1.7$ & = p42\\
29 & kerrbb   & rflag & $1^{\dagger}$ & \\
30 & kerrbb   & lflag & $0^{\dagger}$ & \\
31 & kerrbb   & norm & $1.0^{\dagger}$ & \\
32 & thcomp   & $\tau$ & $0.40^{+0.05}_{-0.06}$ &  \\
33 & thcomp   & $kT_e$ (keV) & $150^{\dagger}$ & \\
34 & thcomp   & cov\_frac & $0.005^{+0.027}_{-0.004}$ &  \\
35 & thcomp   & $z$ & $0.0^{\dagger}$ & \\
36 & kerrbb   & $\eta$ & $0.0^{\dagger}$ & \\
37 & kerrbb   & $a$ & $0.91$ &  = p7\\
38 & kerrbb   & $i$ (deg) & $57.40$  & = p8\\
39 & kerrbb   & $M_{\rm BH}$ ($M_\odot$) & $11.0^{\dagger}$ & \\
40 & kerrbb   & $\dot{M}$ & $0.12^{+0.02}_{-0.02}$ & \\
41 & kerrbb   & $D_{\rm BH}$ (kpc) & $5.4^{\dagger}$ & \\
42 & kerrbb   & $hd$ & $1.7^{\dagger}$ & \\
43 & kerrbb   & rflag & $1^{\dagger}$ & \\
44 & kerrbb   & lflag & $0^{\dagger}$ & \\
45 & kerrbb   & norm & $1.0^{\dagger}$ & \\
46 & constant & factor & $0.73^{+0.02}_{-0.02}$ & LAXPC20 \\
91 & constant & factor & $0.94^{+0.01}_{-0.01}$ & NICER \\
\hline
\end{tabular}
\end{table}
Here, the reflection is modeled using \texttt{xilconv} \citep{Done...2006}, which accounts for features such as the Fe K$\alpha$ line and the Compton hump. Since this reflected radiation originates from regions close to the black hole, it experiences significant relativistic distortion, which is accounted for by the \texttt{relconv} convolution model \citep{Dauser...2010,Dauser...2013}. Physically, the fraction of photons scattering through the corona depends on two factors; covering fraction and optical depth of the corona. The probability that a photon, which enters the corona, scatters at least once is given by (1 - e$^{-\tau}$) \citep{Steiner2009,RybickiANDLightman}. Hence, we include \texttt{constant} which is equal to the covering fraction $\times$ (1 - e$^{-\tau}$), to take care of the number of photons entering the corona and their scattering probability in the corona of optical depth $\tau$. Now, this forms the first branch of our model: \texttt{constant} $\times$ \texttt{relconv} $\times$ \texttt{xilconv} $\times$ \texttt{thcomp} $\times$ \texttt{kerrbb}.

At the same time, not all Comptonized photons are reflected; a significant portion escapes directly to the observer without further interaction. This is accounted for by the second branch, \texttt{thcomp}~$\times$~\texttt{kerrbb}, which models the direct Comptonized continuum along with the underlying thermal disk emission. By separately modeling the reflected and direct components, this model allows for tighter constraints on the estimation of the key parameters, such as black hole spin and disk inclination, while avoiding the degeneracies and assumptions often present in more simplified, coupled models. This model resulted in a statistically acceptable fit, with a chi-square value of 99.79 for 152 degrees of freedom.  We note that a 2\% systematic uncertainty is included in all spectral fits to account for known calibration uncertainties of the instrument. The inclusion of this systematic term naturally reduces the reduced $\chi^2$ values and can lead to values below unity. Although Model~3S does not provide a lower $\chi^2$ than the simpler models, it offers a more self-consistent physical description of the observed reflection features; hence, we continue with Model 3S. To ensure physical consistency, we tie several parameters across different components of the model. Specifically, the spin parameter and inclination angle of the \texttt{kerrbb} component in the second branch are tied to those of \texttt{relconv}, and all parameters of the \texttt{kerrbb} in the first branch are tied to those of the second branch, with its normalization fixed to unity. The optical depth ($\tau$) and coronal temperature (kt$_{\mathrm{e}}$) of \texttt{thcomp} in the first branch are tied to the corresponding parameters of \texttt{thcomp} in the second branch. The covering fraction of \texttt{thcomp} in the first branch is fixed at 1, while in the second branch it is kept free. For \texttt{xilconv}, the rel$\_$refl parameter is set to negative to make sure only the reflected component is returned, and this parameter is allowed to vary during fitting. The fit yielded well-constrained values for key physical parameters, including a black hole spin of $a = 0.91^{+0.03}_{-0.05}$ and a disk inclination angle of $57.40^{+3.13}_{-3.47}$ degrees, assuming a black hole mass of 11\,M$_\odot$ and a source distance of 5.4\,kpc. We observe the value of relative reflection to be large. Within the framework of lamp-post coronal geometry, a high reflection fraction is observed when the corona is located at small heights above the black hole \citep{Miniutti_Fabian_2004, Dauser_2016, Zdziarski_2026}. In this regime, strong gravitational light bending redirects a substantial fraction of Comptonized photons toward the accretion disk, thereby increasing the observed reflected emission relative to the continuum. Furthermore, \cite{Dauser_2014, Dauser_2016} suggested a higher reflection fraction for highly rotating black holes, particularly when the corona height is low. We list the values of all the spectral parameters, varied or fixed, observed after fitting the energy spectrum with model 3S for clarity in Table \ref{Tab: All Spectral Parameters}.

\setlength{\tabcolsep}{10pt} 
\begin{deluxetable}{ccccc}
\label{Table: Spin and Inclination Values for hd 1.5}
\tabletypesize{\footnotesize}
\tablecolumns{5}
\tablewidth{\columnwidth}

\tablecaption{Variation of spin parameter and inclination angle with different combinations of black hole mass and source distance with hardening factor fixed at 1.5. Errors are provided with a 90\% confidence level. \label{tab:spin_incl_grid_hd1.5}}

\tablehead{
\colhead{\textbf{Mass}} & \colhead{\textbf{Dist.}} & \colhead{\textbf{Spin}} & \colhead{\textbf{Incl.}} & \colhead{\textbf{$\chi^2$/dof}} \\
\colhead{($M_\odot$)} & \colhead{(kpc)} & \colhead{($a$)} & \colhead{(deg)} & \colhead{}
}

\startdata
7.26  & 3.3 &  $0.97^{+0.02}_{-0.04}$ &  $58.98^{+5.18}_{-5.74}$ &  98.60/152 \\
7.26  & 5.4 &  $0.87^{+0.02}_{-0.05}$ &  $55.89^{+4.08}_{-1.27}$ &  101.29/152 \\
7.26  & 7.5 &  $<0.78$               &  $55.87^{+10.04}_{-5.13}$ &  103.24/152 \\
\hline
11.00 & 3.3 &  $>0.98$               &  $62.31^{+3.91}_{-1.16}$ &  97.90/152 \\
11.00 & 5.4 &  $0.95^{+0.04}_{-0.04}$ &  $59.77^{+3.55}_{-6.74}$ &  99.10/152 \\
11.00 & 7.5 &  $>0.83$ &  $57.72^{+3.14}_{-2.37}$ &  100.67/152 \\
\hline
16.50 & 3.3 &  $>0.98$               &  $71.32^{+1.53}_{-0.85}$ &  102.56/152 \\
16.50 & 5.4 &  $>0.98$               &  $61.04^{+3.67}_{-1.99}$ &  97.57/152 \\
16.50 & 7.5 &  $>0.93$ &  $58.76^{+3.12}_{-4.89}$ &  98.53/152 \\
\enddata

\end{deluxetable}
\setlength{\tabcolsep}{6pt} 

We perform spectral analysis assuming a black hole mass of 11\,M$_\odot$ and a source distance of 5.4\,kpc. The distance estimate of $5.4^{+2.1}_{-2.1}$\,kpc is reported in \citet{Gaia...2016}, while a separate work by \citet{Campana...2002} place constraints on the black hole mass as a function of distance. Specifically, they suggest that for a maximally spinning black hole, the mass must satisfy $M_{\rm BH} \geq 22 \times d_{10}$\,M$_\odot$, and for a non-spinning black hole, $M_{\rm BH} \geq 3.7 \times d_{10}$\,M$_\odot$, where $d_{10}$ is the distance in units of 10\,kpc.

\begin{figure*}[htbp]
    \centering
    \includegraphics[width=\textwidth]{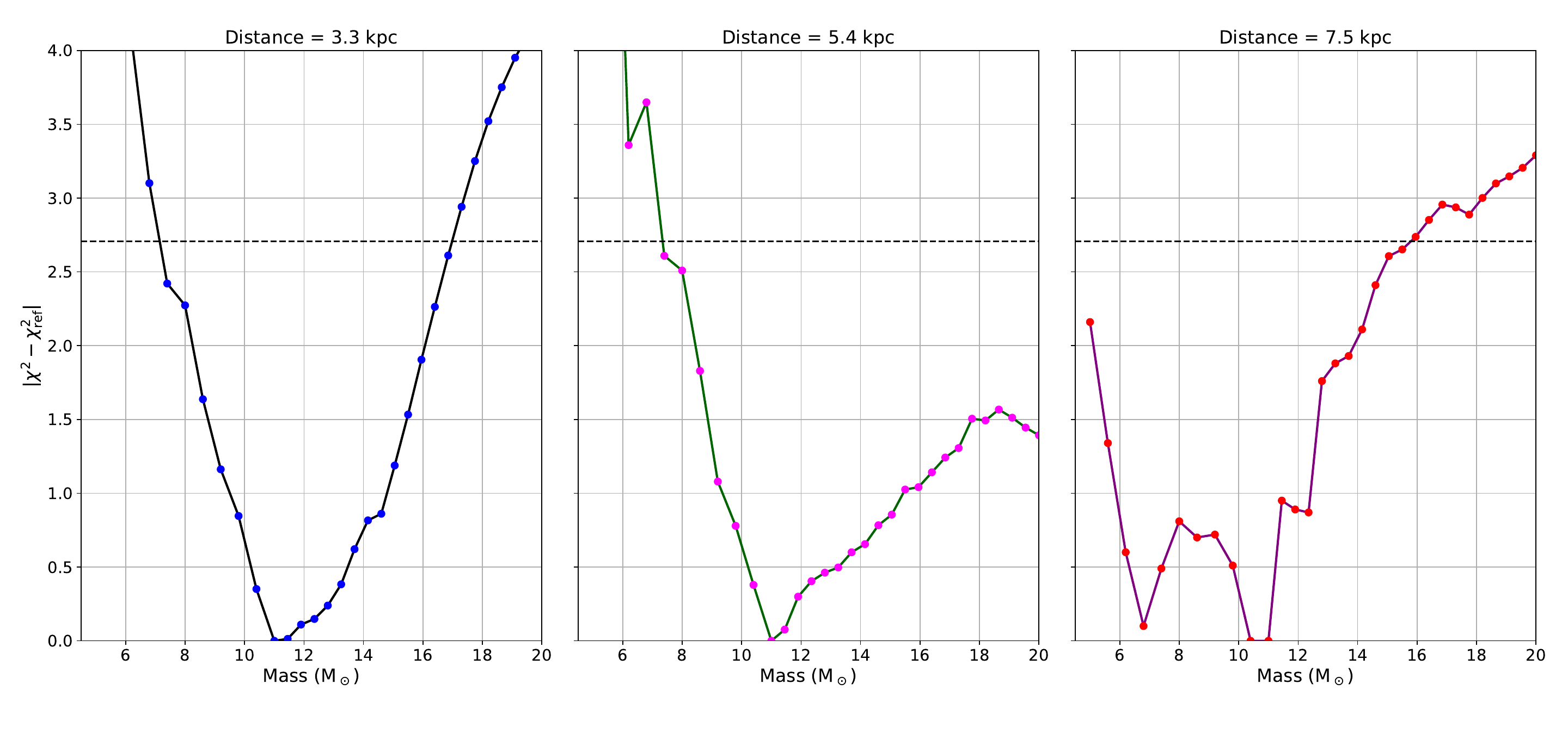}
    \caption{Fit statistic ($\chi^2$) as a function of black hole mass, obtained using the \texttt{steppar} command in \textsc{xspec}, at three assumed source distances: 3.3\,kpc, 5.4\,kpc, and 7.5\,kpc. In each case, the black hole mass was varied between 5 and 20\,M$_\odot$.  The horizontal dashed line represents the 90$\%$ confidence level.}
    \label{fig:steppar_mass}
\end{figure*}

Given the uncertainties in both the mass of the black hole and the distance of the source, we study how changes in the assumed black hole mass affect the overall fit quality of our model. To do this, we use the \texttt{steppar} command in \textsc{XSPEC} to systematically vary the black hole mass between 5\,M$_\odot$ and 20\,M$_\odot$, allowing all other parameters to remain free during the fit. This analysis was repeated at three representative values of the source distance — 3.3\,kpc, 5.4\,kpc, and 7.5\,kpc — which range within the reported values in the literature. Figure \ref{fig:steppar_mass} shows that the estimated black hole mass varies with the assumed source distance. For a distance of 3.3 kpc, the mass is constrained to lie between $\sim$7 and $\sim$17 M$_{\odot}$. When the distance is assumed to be 5.4 kpc, the mass has a lower limit of approximately 7 M$_{\odot}$. In contrast, for a distance of 7.5 kpc, the upper limit of the mass is constrained to about 16 M$_{\odot}$.

Further, we study how variations in the assumed values of black hole mass and the source distance affect the estimation of spin and disk inclination angle values. Specifically, we assume combinations of black hole mass values at 7.26\,M$_\odot$, 11\,M$_\odot$, and 16.5\,M$_\odot$, source distances of 3.3\,kpc, 5.4\,kpc, and 7.5\,kpc, along with spectral hardening factor ($hd$) values of 1.5, 1.7, and 1.9. For each assumed combination, we perform spectral fits using our final model 3S to get the estimation on spin parameter and inclination angle values. This allows us to understand how sensitive the values of spin parameters and inclination angle are to the assumed combinations of the black hole mass, the source distance, and the spectral hardening factor. The results are listed in Table~\ref{Table: Spin and Inclination Values for hd 1.5}, \ref{Table: Spin and Inclination Values for hd 1.7} and \ref{Table: Spin and Inclination Values for hd 1.9}.

\setlength{\tabcolsep}{10pt}
\begin{deluxetable}{ccccc}
\label{Table: Spin and Inclination Values for hd 1.7}
\tabletypesize{\footnotesize}
\tablecolumns{5}
\tablewidth{\columnwidth}

\tablecaption{Variation of spin parameter and inclination angle with different combinations of black hole mass and source distance with hardening factor fixed at 1.7. Errors are provided with a 90\% confidence level. \label{tab:spin_incl_grid_hd1.7}}

\tablehead{
\colhead{Mass} & \colhead{Dist.} & \colhead{Spin} & \colhead{Incl.} & \colhead{$\chi^2$/dof} \\
\colhead{($M_\odot$)} & \colhead{(kpc)} & \colhead{($a$)} & \colhead{(deg)} & \colhead{}
}

\startdata
7.26 & 3.3 &  $0.90^{+0.03}_{-0.04}$ &  $59.99^{+5.69}_{-4.05}$ &  98.40/152 \\
7.26 & 5.4 &  $0.77^{+0.03}_{-0.10}$ &  $54.66^{+5.12}_{-2.54}$ &  102.41/154 \\
7.26 & 7.5 &  $<0.63$               &  $56.03^{+1.75}_{-5.80}$ &  102.95/154 \\
\hline
11.00 & 3.3 &  $>0.96$              &  $58.44^{+6.35}_{-3.24}$ &  97.66/152 \\
11.00 & 5.4 &  $0.91^{+0.03}_{-0.05}$ &  $57.40^{+3.13}_{-3.47}$ &  99.79/152 \\
11.00 & 7.5 &  $0.81^{+0.02}_{-0.08}$ &  $54.97^{+4.16}_{-3.02}$ &  102.47/152 \\
\hline
16.50 & 3.3 &  $>0.98$              &  $65.80^{+1.96}_{-1.02}$ &  99.82/152 \\
16.50 & 5.4 &  $>0.94$              &  $58.12^{+6.60}_{-4.10}$ &  98.17/152 \\
16.50 & 7.5 &  $0.92^{+0.04}_{-0.04}$ &  $58.21^{+3.42}_{-4.50}$ &  99.93/152 \\
\enddata

\end{deluxetable}

\setlength{\tabcolsep}{10pt} 
\begin{deluxetable}{ccccc}
\label{Table: Spin and Inclination Values for hd 1.9}
\tabletypesize{\footnotesize}
\tablecolumns{5}
\tablewidth{\columnwidth}

\tablecaption{Variation of spin parameter and inclination angle with different combinations of black hole mass and source distance with hardening factor fixed at 1.9. Errors are provided with a 90\% confidence level. \label{tab:spin_incl_grid_hd1.9}}

\tablehead{
\colhead{\textbf{Mass}} & \colhead{\textbf{Dist.}} & \colhead{\textbf{Spin}} & \colhead{\textbf{Incl.}} & \colhead{\textbf{$\chi^2$/dof}} \\
\colhead{($M_\odot$)} & \colhead{(kpc)} & \colhead{($a$)} & \colhead{(deg)} & \colhead{}
}

\startdata
7.26  & 3.3 &  $0.87^{+0.04}_{-0.02}$ &  $56.00^{+3.77}_{-2.66}$ &  100.77/152 \\
7.26  & 5.4 &  $<0.71$               &  $55.84^{+25.00}_{-5.75}$ &  103.29/152 \\
7.26  & 7.5 &  $<0.43$               &  $55.43^{+7.63}_{-5.62}$ &  101.24/152 \\
\hline
11.00 & 3.3 &  $0.97^{+0.02}_{-0.05}$ &  $58.27^{+5.43}_{-5.30}$ &  99.02/152 \\
11.00 & 5.4 &  $0.84^{+0.04}_{-0.07}$ &  $56.33^{+4.98}_{-4.30}$ &  101.81/152 \\
11.00 & 7.5 &  $0.70^{+0.04}_{-0.2}$               &  $54.21^{+6.38}_{-2.90}$ &  103.11/152 \\
\hline
16.50 & 3.3 &  $>0.98$               &  $60.72^{+4.16}_{-1.18}$ &  97.45/152 \\
16.50 & 5.4 &  $0.94^{+0.04}_{-0.04}$ &  $59.18^{+3.54}_{-5.85}$ &  99.28/152 \\
16.50 & 7.5 &  $0.87^{+0.04}_{-0.05}$ &  $56.04^{+3.76}_{-2.59}$ &  100.88/152 \\
\enddata

\end{deluxetable}
\setlength{\tabcolsep}{6pt} 

\section{Discussion and Conclusion} \label{sec:Discussion}
In this work, we use simultaneous observations from NICER and AstroSat to study the spectral properties of XTE J2012+381 during its 2022 outburst. We start our spectral analysis by modeling the energy spectrum of the source with phenomenological models. Then, we adopt more physically motivated models that includes thermal disk emission, Comptonization, and relativistic reflection and the final model 

\texttt{constant $\times$ TBabs $\times$ (constant $\times$ relconv $\times$ xilconv $\times$ thcomp $\times$ kerrbb + thcomp $\times$ kerrbb)}

accounts for both the reflected and direct Comptonized components, and provided a statistically good fit (Figure \ref{fig:EnergySpectrum}).

Estimation of both black hole mass and source distance is still not well constrained for this source. \cite{Campana...2002} provided mass estimates that depend strongly on distance, with a suggested lower limit of 22 $\times$ d$_{10}$ M$_\odot$ for a maximally spinning black hole, where d$_{10}$ is the distance of the source in units of 10 kpc. The distance itself is typically taken as $5.4^{+2.1}_{-2.1}$ kpc, based on Gaia parallax measurements (Gaia Collaboration 2016). Given this uncertainty, we first attempt to constrain the black hole mass at different assumed source distances of 3.3, 5.4, and 7.5 kpc using the \texttt{steppar} command in \textsc{XSPEC}. We find that the inferred mass range depends sensitively on the adopted distance. For a distance of 3.3 kpc, the allowed mass spans approximately $\sim$7–17 M${\odot}$. When the distance is fixed at 5.4 kpc, the solutions favor comparatively higher masses, yielding a lower limit of about 7 M${\odot}$. In contrast, assuming a distance of 7.5 kpc restricts the upper bound of the mass to $\sim$16 M$_{\odot}$. This behaviour reflects the degeneracy between mass and distance in spectral modeling, where variations in the assumed distance directly propagate into systematic shifts in the inferred black hole mass.

Next, we study how different combinations of black hole mass (7.26–16.5 M$_\odot$), distance (3.3–7.5 kpc), and spectral hardening factor (1.5-1.9) affect the estimation of the spin parameter and disk inclination angle. Tables~\ref{tab:spin_incl_grid_hd1.5}, \ref{Table: Spin and Inclination Values for hd 1.7}, and \ref{tab:spin_incl_grid_hd1.9} present the dependence of the inferred spin parameter and inclination angle on the assumed black hole mass, source distance, and spectral hardening factor ($hd=1.5$, 1.7, and 1.9, respectively). The $\chi^2$/dof values reported in Tables~\ref{tab:spin_incl_grid_hd1.5}--\ref{tab:spin_incl_grid_hd1.9} remain statistically comparable for all parameter combinations, typically within $\sim 97$--103 for $152$ degrees of freedom. The results clearly demonstrate that the spin estimate is sensitive to the adopted system parameters, while the inclination remains comparatively stable. For $hd=1.5$ (Table~\ref{tab:spin_incl_grid_hd1.5}), the solutions generally favor a rapidly rotating black hole. In particular, for $M=11$ and $16.5\,M_\odot$, the spin is consistently constrained to $a \gtrsim 0.83$, with several cases pegged at $a > 0.98$. Even for the lower mass of $7.26\,M_\odot$, the inferred spin remains high ($a \sim 0.82$--0.99) at smaller distances, although it prefers lower spin values when the distance is increased to 7.5 kpc. It indicates that, for relatively low hardening factors, the spectral fits prefer a high black hole spin.

When the hardening factor is increased to $hd=1.7$ (Table~\ref{tab:spin_incl_grid_hd1.7}), the overall behavior remains similar with prefering spins $\gtrsim 0.68$ for masses $11\,M_\odot$ and $16\,M_\odot$, but with reduced spin values for mass $7.26\,M_\odot$. For $hd=1.9$ (Table~\ref{tab:spin_incl_grid_hd1.9}), this trend becomes more prominent. In the extreme combination of low mass ($7.26\,M_\odot$) and large distance (7.5 kpc), the spin is again constrained only by an upper limit ($a < 0.43$), indicating that low-spin solutions become statistically permissible under this mass-distance assumption. However, for intermediate and higher masses, the spin remains in high regime. For $M=11\,M_\odot$, the spin spans $a \sim 0.68$--0.99, while for $M=16.5\,M_\odot$ it remains $a \gtrsim 0.78$ across all distances taken. Thus, except for the most extreme combination of low mass and large distance, the black hole spin consistently appears to be in high regime.

In contrast to the spin parameter, the inclination angle shows comparatively weak dependence on the assumed system parameters. Across Tables~\ref{tab:spin_incl_grid_hd1.5}--\ref{tab:spin_incl_grid_hd1.9}, the inclination is largely confined to $\sim 50^\circ$-$65^\circ$, with only one notably higher value of $\sim 71^\circ$ obtained for the specific case of $M=16.5\,M_\odot$, $D=3.3$ kpc, and $hd=1.5$ (Table~\ref{tab:spin_incl_grid_hd1.5}). Considering all mass, distance, and hardening factor combinations, a representative inclination range of $i \approx 55^\circ$--$65^\circ$ appears reasonable.

In conclusion, the best-fit model yields a black hole spin of 
$a = 0.91^{+0.03}_{-0.05},(\mathrm{statistical}), ^{+0.09}_{-0.24}\,(\mathrm{systematic})$ 
and an inclination angle of 
$i = 57.40^{+3.13}_{-3.47} ( \mathrm{statistical}), ^{+13.6}_{-7.4}(\mathrm{systematic})$. The statistical uncertainties are quoted with a 90$\%$ confidence level, and the systematic uncertainties are estimated by varying the black hole mass, source distance, and spectral hardening factor within the ranges reported in the literature.

\begin{acknowledgments}
We are grateful for the data from the LAXPC and SXT instruments on board the AstroSat satellite and for the data from the XTI onboard the NICER, which we use in this study. The analysis is carried out using LAXPC software, the SXT pipeline, the nicerl2 pipeline, and HEASoft tools. I, Vaibhav Sharma, appreciate the support and warm hospitality from the Inter-University Centre for Astronomy and Astrophysics (IUCAA) during my visit, which help me greatly in completing this work. I would also like to gratefully acknowledge Divya Rawat (Observatoire astronomique de Strasbourg, France) for her guidance and assistance in learning AstroSat data analysis.
\end{acknowledgments}

\begin{contribution}

All authors contribute equally.


\end{contribution}

%
\facilities{AstroSat (LAXPC and SXT), NICER (XTI)}

\software{
LAXPCSoftware, Heasoft, XSPEC          }


\bibliography{Bibliography}{}

@ARTICLE{Dauser_2016,
       author = {{Dauser}, T. and {Garc{\'\i}a}, J. and {Walton}, D.~J. and {Eikmann}, W. and {Kallman}, T. and {McClintock}, J. and {Wilms}, J.},
        title = "{Normalizing a relativistic model of X-ray reflection. Definition of the reflection fraction and its implementation in relxill}",
      journal = {\aap},
         year = 2016,
        month = may,
       volume = {590},
          eid = {A76},
        pages = {A76},
          doi = {10.1051/0004-6361/201628135},
archivePrefix = {arXiv},
       eprint = {1601.03771},
 primaryClass = {astro-ph.HE},
       adsurl = {https://ui.adsabs.harvard.edu/abs/2016A&A...590A..76D}
}

@ARTICLE{Dauser_2014,
       author = {{Dauser}, T. and {Garcia}, J. and {Parker}, M.~L. and {Fabian}, A.~C. and {Wilms}, J.},
        title = "{The role of the reflection fraction in constraining black hole spin.}",
      journal = {\mnras},
         year = 2014,
        month = oct,
       volume = {444},
        pages = {L100-L104},
          doi = {10.1093/mnrasl/slu125},
archivePrefix = {arXiv},
       eprint = {1408.2347},
 primaryClass = {astro-ph.HE},
       adsurl = {https://ui.adsabs.harvard.edu/abs/2014MNRAS.444L.100D}
}

@ARTICLE{Miniutti_Fabian_2004,
       author = {{Miniutti}, G. and {Fabian}, A.~C.},
        title = "{A light bending model for the X-ray temporal and spectral properties of accreting black holes}",
      journal = {\mnras},
         year = 2004,
        month = apr,
       volume = {349},
       number = {4},
        pages = {1435-1448},
          doi = {10.1111/j.1365-2966.2004.07611.x},
archivePrefix = {arXiv},
       eprint = {astro-ph/0309064},
 primaryClass = {astro-ph},
       adsurl = {https://ui.adsabs.harvard.edu/abs/2004MNRAS.349.1435M}
}

@ARTICLE{Zdziarski_2026,
       author = {{Zdziarski}, Andrzej A. and {Chand}, Swadesh and {Dewangan}, Gulab and {Misra}, Ranjeev and {Szanecki}, Micha{\l} and {You}, Bei and {Parra}, Maxime and {Marcel}, Gr{\'e}goire},
        title = "{The Strong Fe K Line and Spin of the Black Hole X-Ray Binary MAXI J1631─479}",
      journal = {\apjl},
         year = 2026,
        month = feb,
       volume = {998},
       number = {2},
          eid = {L37},
        pages = {L37},
          doi = {10.3847/2041-8213/ae3e8b},
archivePrefix = {arXiv},
       eprint = {2511.03386},
 primaryClass = {astro-ph.HE},
       adsurl = {https://ui.adsabs.harvard.edu/abs/2026ApJ...998L..37Z}}

@ARTICLE{Remillard_2006,
       author = {{Remillard}, Ronald A. and {McClintock}, Jeffrey E.},
        title = "{X-Ray Properties of Black-Hole Binaries}",
      journal = {\araa},
         year = 2006,
        month = sep,
       volume = {44},
       number = {1},
        pages = {49-92},
          doi = {10.1146/annurev.astro.44.051905.092532},
archivePrefix = {arXiv},
       eprint = {astro-ph/0606352},
 primaryClass = {astro-ph},
       adsurl = {https://ui.adsabs.harvard.edu/abs/2006ARA&A..44...49R}
}

@ARTICLE{Done_2007,
       author = {{Done}, Chris and {Gierli{\'n}ski}, Marek and {Kubota}, Aya},
        title = "{Modelling the behaviour of accretion flows in X-ray binaries. Everything you always wanted to know about accretion but were afraid to ask}",
      journal = {\aapr},
         year = 2007,
        month = dec,
       volume = {15},
       number = {1},
        pages = {1-66},
          doi = {10.1007/s00159-007-0006-1},
archivePrefix = {arXiv},
       eprint = {0708.0148},
 primaryClass = {astro-ph},
       adsurl = {https://ui.adsabs.harvard.edu/abs/2007A&ARv..15....1D}
}

@INPROCEEDINGS{Singh_2014,
       author = {{Singh}, Kulinder Pal and {Tandon}, S. N. and {Agrawal}, P. C. and {Antia}, H. M. and {Manchanda}, R. K. and {Yadav}, J. S. and {Seetha}, S. and {Ramadevi}, M. C. and {Rao}, A. R. and {Bhattacharya}, D. and {Paul}, B. and {Sreekumar}, P. and {Bhattacharyya}, S. and {Stewart}, G. C. and {Hutchings}, J. and {Annapurni}, S. A. and {Ghosh}, S. K. and {Murthy}, J. and {Pati}, A. and {Rao}, N. K. and {Stalin}, C. S. and {Girish}, V. and {Sankarasubramanian}, K. and {Vadawale}, S. and {Bhalerao}, V. B. and {Dewangan}, G. C. and {Dedhia}, D. K. and {Hingar}, M. K. and {Katoch}, T. B. and {Kothare}, A. T. and {Mirza}, I. and {Mukerjee}, K. and {Shah}, H. and {Shah}, P. and {Mohan}, R. and {Sangal}, A. K. and {Nagabhusana}, S. and {Sriram}, S. and {Malkar}, J. P. and {Sreekumar}, S. and {Abbey}, A. F. and {Hansford}, G. M. and {Beardmore}, A. P. and {Sharma}, M. R. and {Murthy}, S. and {Kulkarni}, R. and {Meena}, G. and {Babu}, V. C. and {Postma}, J.},
        title = "{ASTROSAT mission}",
    booktitle = {Space Telescopes and Instrumentation 2014: Ultraviolet to Gamma Ray},
         year = 2014,
       editor = {{Takahashi}, Tadayuki and {den Herder}, Jan-Willem A. and {Bautz}, Mark},
       series = {Society of Photo-Optical Instrumentation Engineers (SPIE) Conference Series},
       volume = {9144},
        month = jul,
          eid = {91441S},
        pages = {91441S},
          doi = {10.1117/12.2062667},
       adsurl = {https://ui.adsabs.harvard.edu/abs/2014SPIE.9144E..1SS}
}

@ARTICLE{Steiner2009,
       author = {{Steiner}, James F. and {Narayan}, Ramesh and {McClintock}, Jeffrey E. and {Ebisawa}, Ken},
        title = "{A Simple Comptonization Model}",
      journal = {\pasp},
         year = 2009,
        month = nov,
       volume = {121},
       number = {885},
        pages = {1279},
          doi = {10.1086/648535},
archivePrefix = {arXiv},
       eprint = {0810.1758},
 primaryClass = {astro-ph},
       adsurl = {https://ui.adsabs.harvard.edu/abs/2009PASP..121.1279S}
}

@BOOK{RybickiANDLightman,
       author = {{Rybicki}, George B. and {Lightman}, Alan P.},
        title = "{Radiative processes in astrophysics}",
         year = 1979,
       adsurl = {https://ui.adsabs.harvard.edu/abs/1979rpa..book.....R}
}

@ARTICLE{Dauser...2013,
       author = {{Dauser}, T. and {Garcia}, J. and {Wilms}, J. and {B{\"o}ck}, M. and {Brenneman}, L.~W. and {Falanga}, M. and {Fukumura}, K. and {Reynolds}, C.~S.},
        title = "{Irradiation of an accretion disc by a jet: general properties and implications for spin measurements of black holes}",
      journal = {\mnras},
         year = 2013,
        month = apr,
       volume = {430},
       number = {3},
        pages = {1694-1708},
          doi = {10.1093/mnras/sts710},
archivePrefix = {arXiv},
       eprint = {1301.4922},
 primaryClass = {astro-ph.HE},
       adsurl = {https://ui.adsabs.harvard.edu/abs/2013MNRAS.430.1694D}
}

@ARTICLE{Dauser...2010,
       author = {{Dauser}, T. and {Wilms}, J. and {Reynolds}, C.~S. and {Brenneman}, L.~W.},
        title = "{Broad emission lines for a negatively spinning black hole}",
      journal = {\mnras},
         year = 2010,
        month = dec,
       volume = {409},
       number = {4},
        pages = {1534-1540},
          doi = {10.1111/j.1365-2966.2010.17393.x},
archivePrefix = {arXiv},
       eprint = {1007.4937},
 primaryClass = {astro-ph.HE},
       adsurl = {https://ui.adsabs.harvard.edu/abs/2010MNRAS.409.1534D}
}

@ARTICLE{Done...2006,
       author = {{Done}, Chris and {Gierli{\'n}ski}, Marek},
        title = "{Truncated disc versus extremely broad iron line in XTE J1650-500}",
      journal = {\mnras},
         year = 2006,
        month = apr,
       volume = {367},
       number = {2},
        pages = {659-668},
          doi = {10.1111/j.1365-2966.2005.09968.x},
archivePrefix = {arXiv},
       eprint = {astro-ph/0510614},
 primaryClass = {astro-ph},
       adsurl = {https://ui.adsabs.harvard.edu/abs/2006MNRAS.367..659D}
}

@ARTICLE{Li...2005,
       author = {{Li}, Li-Xin and {Zimmerman}, Erik R. and {Narayan}, Ramesh and {McClintock}, Jeffrey E.},
        title = "{Multitemperature Blackbody Spectrum of a Thin Accretion Disk around a Kerr Black Hole: Model Computations and Comparison with Observations}",
      journal = {\apjs},
         year = 2005,
        month = apr,
       volume = {157},
       number = {2},
        pages = {335-370},
          doi = {10.1086/428089},
archivePrefix = {arXiv},
       eprint = {astro-ph/0411583},
 primaryClass = {astro-ph},
       adsurl = {https://ui.adsabs.harvard.edu/abs/2005ApJS..157..335L}
}

@article{Zdziarski2020,
  author = {Zdziarski, A. A. and Szanecki, M. and Poutanen, J. and Gierliński, M. and Biernacki, P.},
  title = {Spectral and temporal properties of Compton scattering by mildly relativistic thermal electrons},
  journal = {Monthly Notices of the Royal Astronomical Society},
  volume = {492},
  pages = {5234--5246},
  year = {2020},
  doi = {10.1093/mnras/staa159}
}

@article{Wilms2000,
  author = {Wilms, J. and Allen, A. and McCray, R.},
  title = {On the Absorption of X-Rays in the Interstellar Medium},
  journal = {The Astrophysical Journal},
  volume = {542},
  number = {2},
  pages = {914--924},
  year = {2000},
  doi = {10.1086/317016}
}

@article{Garcia2014,
  author = {García, J. and Dauser, T. and Lohfink, A. and Kallman, T. R. and Steiner, J. F. and McClintock, J. E. and Brenneman, L. W. and Wilms, J. and Eikmann, W. and Reynolds, C. S. and Tombesi, F.},
  title = {Improved Reflection Models of Black-Hole Accretion Disks: Treating the Angular Distribution of X-rays},
  journal = {The Astrophysical Journal},
  volume = {782},
  number = {2},
  pages = {76},
  year = {2014},
  doi = {10.1088/0004-637X/782/2/76}
}

@article{Mitsuda...1984,
  author    = {Mitsuda, K. and Inoue, H. and Koyama, K. and others},
  title     = {Energy spectra of low-mass binary X-ray sources observed from Tenma},
  journal   = {Publications of the Astronomical Society of Japan},
  year      = {1984},
  volume    = {36},
  pages     = {741},
  URL       = {https://ui.adsabs.harvard.edu/abs/1984PASJ...36..741M/abstract}
}

@ARTICLE{Raj...2024,
       author = {{Kumar}, Raj},
        title = "{Black hole spin estimation of XTE J2012+381 using simultaneous observations of Swift/XRT and NuSTAR}",
      journal = {\mnras},
         year = 2024,
        month = aug,
       volume = {532},
       number = {2},
        pages = {2635-2643},
          doi = {10.1093/mnras/stae1683},
archivePrefix = {arXiv},
       eprint = {2407.07362},
 primaryClass = {astro-ph.HE},
       adsurl = {https://ui.adsabs.harvard.edu/abs/2024MNRAS.532.2635K}
}

@ARTICLE{Draghis...2023a,
       author = {{Draghis}, Paul A. and {Miller}, Jon M. and {Brumback}, McKinley C. and {Fabian}, Andrew C. and {Tomsick}, John A. and {Zoghbi}, Abderahmen},
        title = "{An Extreme Black Hole in the Recurrent X-Ray Transient XTE J2012+381}",
      journal = {\apj},
         year = 2023,
        month = sep,
       volume = {954},
       number = {1},
          eid = {62},
        pages = {62},
          doi = {10.3847/1538-4357/ace7b3},
archivePrefix = {arXiv},
       eprint = {2307.06988},
 primaryClass = {astro-ph.HE},
       adsurl = {https://ui.adsabs.harvard.edu/abs/2023ApJ...954...62D}
}

@ARTICLE{NICER...ATel,
       author = {{Iwakiri}, W. and {Shidatsu}, M. and {Gendreau}, K.~C. and {Arzoumanian}, Z. and {Sanna}, A. and {Mihara}, T.},
        title = "{NICER observations of new outburst from the black hole candidate XTE J2012+381}",
      journal = {The Astronomer's Telegram},
         year = 2022,
        month = dec,
       volume = {15829},
        pages = {1},
       adsurl = {https://ui.adsabs.harvard.edu/abs/2022ATel15829....1I}
}

@ARTICLE{SWIFT...ATel,
       author = {{Kennea}, J.~A.},
        title = "{XTE J2012+381: Swift confirmation of new outburst}",
      journal = {The Astronomer's Telegram},
         year = 2022,
        month = dec,
       volume = {15827},
        pages = {1},
       adsurl = {https://ui.adsabs.harvard.edu/abs/2022ATel15827....1K}
}

@ARTICLE{MAXI...ATel,
       author = {{Kawamuro}, T. and {Negoro}, H. and {Nakajima}, M. and {Kobayashi}, K. and {Tanaka}, M. and {Soejima}, Y. and {Mihara}, T. and {Yamada}, S. and {Tamagawa}, T. and {Matsuoka}, M. and {Sakamoto}, T. and {Serino}, M. and {Sugita}, S. and {Hiramatsu}, H. and {Nishikawa}, H. and {Yoshida}, A. and {Tsuboi}, Y. and {Kohara}, J. and {Urabe}, S. and {Nawa}, S. and {Nemoto}, N. and {Iwakiri}, W. and {Shidatsu}, M. and {Iwasaki}, M. and {Kawai}, N. and {Niwano}, M. and {Hosokawa}, R. and {Imai}, Y. and {Ito}, N. and {Takamatsu}, Y. and {Nakahira}, S. and {Ueno}, S. and {Tomida}, H. and {Ishikawa}, M. and {Kurihara}, T. and {Ueda}, Y. and {Ogawa}, S. and {Setoguchi}, K. and {Yoshitake}, T. and {Inaba}, K. and {Yamauchi}, M. and {Sato}, T. and {Hatsuda}, R. and {Fukuoka}, R. and {Hagiwara}, Y. and {Umeki}, Y. and {Yamaoka}, K. and {Kawakubo}, Y. and {Sugizaki}, M.},
        title = "{MAXI/GSC detection of X-ray emission probably from the black hole candidate XTE J2012+381}",
      journal = {The Astronomer's Telegram},
         year = 2022,
        month = dec,
       volume = {15826},
        pages = {1},
       adsurl = {https://ui.adsabs.harvard.edu/abs/2022ATel15826....1K}
}

@ARTICLE{Campana...2002,
       author = {{Campana}, S. and {Stella}, L. and {Belloni}, T. and {Israel}, G.~L. and {Santangelo}, A. and {Frontera}, F. and {Orlandini}, M. and {Dal Fiume}, D.},
        title = "{The 1998 outburst of the X-ray transient XTE J2012+381 as observed with BeppoSAX}",
      journal = {\aap},
         year = 2002,
        month = mar,
       volume = {384},
        pages = {163-170},
          doi = {10.1051/0004-6361:20020012},
archivePrefix = {arXiv},
       eprint = {astro-ph/0112485},
 primaryClass = {astro-ph},
       adsurl = {https://ui.adsabs.harvard.edu/abs/2002A&A...384..163C}
}

@ARTICLE{Vasiliev...2000,
       author = {{Vasiliev}, L. and {Trudolyubov}, S. and {Revnivtsev}, M.},
        title = "{RXTE observations of XTE J2012+381 during its 1998 outburst}",
      journal = {\aap},
         year = 2000,
        month = oct,
       volume = {362},
        pages = {L53-L56},
          doi = {10.48550/arXiv.astro-ph/0008176},
archivePrefix = {arXiv},
       eprint = {astro-ph/0008176},
 primaryClass = {astro-ph},
       adsurl = {https://ui.adsabs.harvard.edu/abs/2000A&A...362L..53V}
}

@ARTICLE{Fishbach...2022,
       author = {{Fishbach}, Maya and {Kalogera}, Vicky},
        title = "{Apples and Oranges: Comparing Black Holes in X-Ray Binaries and Gravitational-wave Sources}",
      journal = {\apjl},
         year = 2022,
        month = apr,
       volume = {929},
       number = {2},
          eid = {L26},
        pages = {L26},
          doi = {10.3847/2041-8213/ac64a5},
archivePrefix = {arXiv},
       eprint = {2111.02935},
 primaryClass = {astro-ph.HE},
       adsurl = {https://ui.adsabs.harvard.edu/abs/2022ApJ...929L..26F}
}

@ARTICLE{Draghis...2023b,
       author = {{Draghis}, Paul A. and {Miller}, Jon M. and {Zoghbi}, Abderahmen and {Reynolds}, Mark and {Costantini}, Elisa and {Gallo}, Luigi C. and {Tomsick}, John A.},
        title = "{A Systematic View of Ten New Black Hole Spins}",
      journal = {\apj},
         year = 2023,
        month = mar,
       volume = {946},
       number = {1},
          eid = {19},
        pages = {19},
          doi = {10.3847/1538-4357/acafe7},
archivePrefix = {arXiv},
       eprint = {2210.02479},
 primaryClass = {astro-ph.HE},
       adsurl = {https://ui.adsabs.harvard.edu/abs/2023ApJ...946...19D}
}

@ARTICLE{Gou...2009,
       author = {{Gou}, Lijun and {McClintock}, Jeffrey E. and {Liu}, Jifeng and {Narayan}, Ramesh and {Steiner}, James F. and {Remillard}, Ronald A. and {Orosz}, Jerome A. and {Davis}, Shane W. and {Ebisawa}, Ken and {Schlegel}, Eric M.},
        title = "{A Determination of the Spin of the Black Hole Primary in LMC X-1}",
      journal = {\apj},
         year = 2009,
        month = aug,
       volume = {701},
       number = {2},
        pages = {1076-1090},
          doi = {10.1088/0004-637X/701/2/1076},
archivePrefix = {arXiv},
       eprint = {0901.0920},
 primaryClass = {astro-ph.HE},
       adsurl = {https://ui.adsabs.harvard.edu/abs/2009ApJ...701.1076G}
}

@ARTICLE{Feng...2023,
       author = {{Feng}, Ye and {Steiner}, James F. and {Ramirez}, Santiago Ubach and {Gou}, Lijun},
        title = "{Using X-ray continuum-fitting to estimate the spin of MAXI J1305-704}",
      journal = {\mnras},
         year = 2023,
        month = apr,
       volume = {520},
       number = {4},
        pages = {5803-5816},
          doi = {10.1093/mnras/stad442},
archivePrefix = {arXiv},
       eprint = {2212.04653},
 primaryClass = {astro-ph.HE},
       adsurl = {https://ui.adsabs.harvard.edu/abs/2023MNRAS.520.5803F}
}

@ARTICLE{Fabian...2000,
       author = {{Fabian}, A.~C. and {Iwasawa}, K. and {Reynolds}, C.~S. and {Young}, A.~J.},
        title = "{Broad Iron Lines in Active Galactic Nuclei}",
      journal = {\pasp},
         year = 2000,
        month = sep,
       volume = {112},
       number = {775},
        pages = {1145-1161},
          doi = {10.1086/316610},
archivePrefix = {arXiv},
       eprint = {astro-ph/0004366},
 primaryClass = {astro-ph},
       adsurl = {https://ui.adsabs.harvard.edu/abs/2000PASP..112.1145F}
}

@ARTICLE{Brenneman...2006,
       author = {{Brenneman}, Laura W. and {Reynolds}, Christopher S.},
        title = "{Constraining Black Hole Spin via X-Ray Spectroscopy}",
      journal = {\apj},
         year = 2006,
        month = dec,
       volume = {652},
       number = {2},
        pages = {1028-1043},
          doi = {10.1086/508146},
archivePrefix = {arXiv},
       eprint = {astro-ph/0608502},
 primaryClass = {astro-ph},
       adsurl = {https://ui.adsabs.harvard.edu/abs/2006ApJ...652.1028B}
}

@ARTICLE{Miller...2007,
       author = {{Miller}, J.~M.},
        title = "{Relativistic X-Ray Lines from the Inner Accretion Disks Around Black Holes}",
      journal = {\araa},
         year = 2007,
        month = sep,
       volume = {45},
       number = {1},
        pages = {441-479},
          doi = {10.1146/annurev.astro.45.051806.110555},
archivePrefix = {arXiv},
       eprint = {0705.0540},
 primaryClass = {astro-ph},
       adsurl = {https://ui.adsabs.harvard.edu/abs/2007ARA&A..45..441M}
}

@ARTICLE{Miller...2010,
       author = {{Miller}, J.~M. and {D'A{\`\i}}, A. and {Bautz}, M.~W. and {Bhattacharyya}, S. and {Burrows}, D.~N. and {Cackett}, E.~M. and {Fabian}, A.~C. and {Freyberg}, M.~J. and {Haberl}, F. and {Kennea}, J. and {Nowak}, M.~A. and {Reis}, R.~C. and {Strohmayer}, T.~E. and {Tsujimoto}, M.},
        title = "{On Relativistic Disk Spectroscopy in Compact Objects with X-ray CCD Cameras}",
      journal = {\apj},
         year = 2010,
        month = dec,
       volume = {724},
       number = {2},
        pages = {1441-1455},
          doi = {10.1088/0004-637X/724/2/1441},
archivePrefix = {arXiv},
       eprint = {1009.4391},
 primaryClass = {astro-ph.HE},
       adsurl = {https://ui.adsabs.harvard.edu/abs/2010ApJ...724.1441M}
}

@ARTICLE{Draghis...2020,
       author = {{Draghis}, Paul A. and {Miller}, Jon M. and {Cackett}, Edward M. and {Kammoun}, Elias S. and {Reynolds}, Mark T. and {Tomsick}, John A. and {Zoghbi}, Abderahmen},
        title = "{A New Spin on an Old Black Hole: NuSTAR Spectroscopy of EXO 1846-031}",
      journal = {\apj},
         year = 2020,
        month = sep,
       volume = {900},
       number = {1},
          eid = {78},
        pages = {78},
          doi = {10.3847/1538-4357/aba2ec},
archivePrefix = {arXiv},
       eprint = {2007.04324},
 primaryClass = {astro-ph.HE},
       adsurl = {https://ui.adsabs.harvard.edu/abs/2020ApJ...900...78D}
}

@ARTICLE{Remillard...1998,
       author = {{Remillard}, R. and {Levine}, A. and {Wood}, A. and {Wagner}, R.~M. and {Starrfield}, S. and {Shrader}, C. and {Bowell}, E. and {Skiff}, B. and {Koehn}, B.},
        title = "{XTE J2012+381}",
      journal = {\iaucirc},
         year = 1998,
        month = may,
       volume = {6920},
        pages = {1},
       adsurl = {https://ui.adsabs.harvard.edu/abs/1998IAUC.6920....1R}
}

@ARTICLE{White...1998,
       author = {{White}, N.~E. and {Ueda}, Y. and {Dotani}, T. and {Nagase}, F.},
        title = "{XTE J2012+381}",
      journal = {\iaucirc},
         year = 1998,
        month = jun,
       volume = {6927},
        pages = {2},
       adsurl = {https://ui.adsabs.harvard.edu/abs/1998IAUC.6927....2W}
}

@ARTICLE{Gaia...2016,
       author = {{Gaia Collaboration} and {Prusti}, T. and {de Bruijne}, J.~H.~J. and {Brown}, A.~G.~A. and {Vallenari}, A. and {Babusiaux}, C. and {Bailer-Jones}, C.~A.~L. and {Bastian}, U. and {Biermann}, M. and {Evans}, D.~W. and {Eyer}, L. and {Jansen}, F. and {Jordi}, C. and {Klioner}, S.~A. and {Lammers}, U. and {Lindegren}, L. and {Luri}, X. and {Mignard}, F. and {Milligan}, D.~J. and {Panem}, C. and {Poinsignon}, V. and {Pourbaix}, D. and {Randich}, S. and {Sarri}, G. and {Sartoretti}, P. and {Siddiqui}, H.~I. and {Soubiran}, C. and {Valette}, V. and {van Leeuwen}, F. and {Walton}, N.~A. and {Aerts}, C. and {Arenou}, F. and {Cropper}, M. and {Drimmel}, R. and {H{\o}g}, E. and {Katz}, D. and {Lattanzi}, M.~G. and {O'Mullane}, W. and {Grebel}, E.~K. and {Holland}, A.~D. and {Huc}, C. and {Passot}, X. and {Bramante}, L. and {Cacciari}, C. and {Casta{\~n}eda}, J. and {Chaoul}, L. and {Cheek}, N. and {De Angeli}, F. and {Fabricius}, C. and {Guerra}, R. and {Hern{\'a}ndez}, J. and {Jean-Antoine-Piccolo}, A. and {Masana}, E. and {Messineo}, R. and {Mowlavi}, N. and {Nienartowicz}, K. and {Ord{\'o}{\~n}ez-Blanco}, D. and {Panuzzo}, P. and {Portell}, J. and {Richards}, P.~J. and {Riello}, M. and {Seabroke}, G.~M. and {Tanga}, P. and {Th{\'e}venin}, F. and {Torra}, J. and {Els}, S.~G. and {Gracia-Abril}, G. and {Comoretto}, G. and {Garcia-Reinaldos}, M. and {Lock}, T. and {Mercier}, E. and {Altmann}, M. and {Andrae}, R. and {Astraatmadja}, T.~L. and {Bellas-Velidis}, I. and {Benson}, K. and {Berthier}, J. and {Blomme}, R. and {Busso}, G. and {Carry}, B. and {Cellino}, A. and {Clementini}, G. and {Cowell}, S. and {Creevey}, O. and {Cuypers}, J. and {Davidson}, M. and {De Ridder}, J. and {de Torres}, A. and {Delchambre}, L. and {Dell'Oro}, A. and {Ducourant}, C. and {Fr{\'e}mat}, Y. and {Garc{\'\i}a-Torres}, M. and {Gosset}, E. and {Halbwachs}, J. -L. and {Hambly}, N.~C. and {Harrison}, D.~L. and {Hauser}, M. and {Hestroffer}, D. and {Hodgkin}, S.~T. and {Huckle}, H.~E. and {Hutton}, A. and {Jasniewicz}, G. and {Jordan}, S. and {Kontizas}, M. and {Korn}, A.~J. and {Lanzafame}, A.~C. and {Manteiga}, M. and {Moitinho}, A. and {Muinonen}, K. and {Osinde}, J. and {Pancino}, E. and {Pauwels}, T. and {Petit}, J. -M. and {Recio-Blanco}, A. and {Robin}, A.~C. and {Sarro}, L.~M. and {Siopis}, C. and {Smith}, M. and {Smith}, K.~W. and {Sozzetti}, A. and {Thuillot}, W. and {van Reeven}, W. and {Viala}, Y. and {Abbas}, U. and {Abreu Aramburu}, A. and {Accart}, S. and {Aguado}, J.~J. and {Allan}, P.~M. and {Allasia}, W. and {Altavilla}, G. and {{\'A}lvarez}, M.~A. and {Alves}, J. and {Anderson}, R.~I. and {Andrei}, A.~H. and {Anglada Varela}, E. and {Antiche}, E. and {Antoja}, T. and {Ant{\'o}n}, S. and {Arcay}, B. and {Atzei}, A. and {Ayache}, L. and {Bach}, N. and {Baker}, S.~G. and {Balaguer-N{\'u}{\~n}ez}, L. and {Barache}, C. and {Barata}, C. and {Barbier}, A. and {Barblan}, F. and {Baroni}, M. and {Barrado y Navascu{\'e}s}, D. and {Barros}, M. and {Barstow}, M.~A. and {Becciani}, U. and {Bellazzini}, M. and {Bellei}, G. and {Bello Garc{\'\i}a}, A. and {Belokurov}, V. and {Bendjoya}, P. and {Berihuete}, A. and {Bianchi}, L. and {Bienaym{\'e}}, O. and {Billebaud}, F. and {Blagorodnova}, N. and {Blanco-Cuaresma}, S. and {Boch}, T. and {Bombrun}, A. and {Borrachero}, R. and {Bouquillon}, S. and {Bourda}, G. and {Bouy}, H. and {Bragaglia}, A. and {Breddels}, M.~A. and {Brouillet}, N. and {Br{\"u}semeister}, T. and {Bucciarelli}, B. and {Budnik}, F. and {Burgess}, P. and {Burgon}, R. and {Burlacu}, A. and {Busonero}, D. and {Buzzi}, R. and {Caffau}, E. and {Cambras}, J. and {Campbell}, H. and {Cancelliere}, R. and {Cantat-Gaudin}, T. and {Carlucci}, T. and {Carrasco}, J.~M. and {Castellani}, M. and {Charlot}, P. and {Charnas}, J. and {Charvet}, P. and {Chassat}, F. and {Chiavassa}, A. and {Clotet}, M. and {Cocozza}, G. and {Collins}, R.~S. and {Collins}, P. and {Costigan}, G.},
        title = "{The Gaia mission}",
      journal = {\aap},
         year = 2016,
        month = nov,
       volume = {595},
          eid = {A1},
        pages = {A1},
          doi = {10.1051/0004-6361/201629272},
archivePrefix = {arXiv},
       eprint = {1609.04153},
 primaryClass = {astro-ph.IM},
       adsurl = {https://ui.adsabs.harvard.edu/abs/2016A&A...595A...1G}
}

@ARTICLE{Zhang...1997,
       author = {{Zhang}, S.~N. and {Cui}, Wei and {Chen}, Wan},
        title = "{Black Hole Spin in X-Ray Binaries: Observational Consequences}",
      journal = {\apjl},
         year = 1997,
        month = jun,
       volume = {482},
       number = {2},
        pages = {L155-L158},
          doi = {10.1086/310705},
archivePrefix = {arXiv},
       eprint = {astro-ph/9704072},
 primaryClass = {astro-ph},
       adsurl = {https://ui.adsabs.harvard.edu/abs/1997ApJ...482L.155Z}
}

@ARTICLE{Tanaka...1995,
       author = {{Tanaka}, Y. and {Nandra}, K. and {Fabian}, A.~C. and {Inoue}, H. and {Otani}, C. and {Dotani}, T. and {Hayashida}, K. and {Iwasawa}, K. and {Kii}, T. and {Kunieda}, H. and {Makino}, F. and {Matsuoka}, M.},
        title = "{Gravitationally redshifted emission implying an accretion disk and massive black hole in the active galaxy MCG-6-30-15}",
      journal = {\nat},
         year = 1995,
        month = jun,
       volume = {375},
       number = {6533},
        pages = {659-661},
          doi = {10.1038/375659a0},
       adsurl = {https://ui.adsabs.harvard.edu/abs/1995Natur.375..659T}
}

@ARTICLE{Fabian...1989,
       author = {{Fabian}, A.~C. and {Rees}, M.~J. and {Stella}, L. and {White}, N.~E.},
        title = "{X-ray fluorescence from the inner disc in Cygnus X-1.}",
      journal = {\mnras},
         year = 1989,
        month = may,
       volume = {238},
        pages = {729-736},
          doi = {10.1093/mnras/238.3.729},
       adsurl = {https://ui.adsabs.harvard.edu/abs/1989MNRAS.238..729F}
}

@INPROCEEDINGS{Yadav...2016,
       author = {{Yadav}, J.~S. and {Agrawal}, P.~C. and {Antia}, H.~M. and {Chauhan}, Jai Verdhan and {Dedhia}, Dhiraj and {Katoch}, Tilak and {Madhwani}, P. and {Manchanda}, R.~K. and {Misra}, Ranjeev and {Pahari}, Mayukh and {Paul}, B. and {Shah}, Parag},
        title = "{Large Area X-ray Proportional Counter (LAXPC) instrument onboard ASTROSAT}",
    booktitle = {Space Telescopes and Instrumentation 2016: Ultraviolet to Gamma Ray},
         year = 2016,
       editor = {{den Herder}, Jan-Willem A. and {Takahashi}, Tadayuki and {Bautz}, Marshall},
       series = {Society of Photo-Optical Instrumentation Engineers (SPIE) Conference Series},
       volume = {9905},
        month = jul,
          eid = {99051D},
        pages = {99051D},
          doi = {10.1117/12.2231857},
       adsurl = {https://ui.adsabs.harvard.edu/abs/2016SPIE.9905E..1DY}
}

@ARTICLE{Agarwal...2017,
       author = {{Agrawal}, P.~C. and {Yadav}, J.~S. and {Antia}, H.~M. and {Dedhia}, Dhiraj and {Shah}, P. and {Chauhan}, Jai Verdhan and {Manchanda}, R.~K. and {Chitnis}, V.~R. and {Gujar}, V.~M. and {Katoch}, Tilak and {Kurhade}, V.~N. and {Madhwani}, P. and {Manojkumar}, T.~K. and {Nikam}, V.~A. and {Pandya}, A.~S. and {Parmar}, J.~V. and {Pawar}, D.~M. and {Roy}, Jayashree and {Paul}, B. and {Pahari}, Mayukh and {Misra}, Ranjeev and {Ravichandran}, M.~H. and {Anilkumar}, K. and {Joseph}, C.~C. and {Navalgund}, K.~H. and {Pandiyan}, R. and {Sarma}, K.~S. and {Subbarao}, K.},
        title = "{Large Area X-Ray Proportional Counter (LAXPC) Instrument on AstroSat and Some Preliminary Results from its Performance in the Orbit}",
      journal = {Journal of Astrophysics and Astronomy},
         year = 2017,
        month = jun,
       volume = {38},
       number = {2},
          eid = {30},
        pages = {30},
          doi = {10.1007/s12036-017-9451-z},
archivePrefix = {arXiv},
       eprint = {1705.06446},
 primaryClass = {astro-ph.IM},
       adsurl = {https://ui.adsabs.harvard.edu/abs/2017JApA...38...30A}
}

@ARTICLE{Antia...2017,
       author = {{Antia}, H.~M. and {Yadav}, J.~S. and {Agrawal}, P.~C. and {Verdhan Chauhan}, Jai and {Manchanda}, R.~K. and {Chitnis}, Varsha and {Paul}, Biswajit and {Dedhia}, Dhiraj and {Shah}, Parag and {Gujar}, V.~M. and {Katoch}, Tilak and {Kurhade}, V.~N. and {Madhwani}, Pankaj and {Manojkumar}, T.~K. and {Nikam}, V.~A. and {Pandya}, A.~S. and {Parmar}, J.~V. and {Pawar}, D.~M. and {Pahari}, Mayukh and {Misra}, Ranjeev and {Navalgund}, K.~H. and {Pandiyan}, R. and {Sharma}, K.~S. and {Subbarao}, K.},
        title = "{Calibration of the Large Area X-Ray Proportional Counter (LAXPC) Instrument on board AstroSat}",
      journal = {\apjs},
         year = 2017,
        month = jul,
       volume = {231},
       number = {1},
          eid = {10},
        pages = {10},
          doi = {10.3847/1538-4365/aa7a0e},
archivePrefix = {arXiv},
       eprint = {1702.08624},
 primaryClass = {astro-ph.IM},
       adsurl = {https://ui.adsabs.harvard.edu/abs/2017ApJS..231...10A}
}

@INPROCEEDINGS{Singh...2016,
       author = {{Singh}, Kulinder Pal and {Stewart}, Gordon C. and {Chandra}, Sunil and {Mukerjee}, Kallol and {Kotak}, Sanket and {Beardmore}, Andy P. and {Chitnis}, Varsha and {Dewangan}, Gulab C. and {Bhattacharyya}, Sudip and {Mirza}, Irfan and {Kamble}, Nilima and {Navalkar}, Vinita and {Shah}, Harshit and {Vishwakarma}, S. and {Koyande}, J.},
        title = "{In-orbit performance of SXT aboard AstroSat}",
    booktitle = {Space Telescopes and Instrumentation 2016: Ultraviolet to Gamma Ray},
         year = 2016,
       editor = {{den Herder}, Jan-Willem A. and {Takahashi}, Tadayuki and {Bautz}, Marshall},
       series = {Society of Photo-Optical Instrumentation Engineers (SPIE) Conference Series},
       volume = {9905},
        month = jul,
          eid = {99051E},
        pages = {99051E},
          doi = {10.1117/12.2235309},
       adsurl = {https://ui.adsabs.harvard.edu/abs/2016SPIE.9905E..1ES}
}

@ARTICLE{Singh...2017,
       author = {{Singh}, K.~P. and {Stewart}, G.~C. and {Westergaard}, N.~J. and {Bhattacharayya}, S. and {Chandra}, S. and {Chitnis}, V.~R. and {Dewangan}, G.~C. and {Kothare}, A.~T. and {Mirza}, I.~M. and {Mukerjee}, K. and {Navalkar}, V. and {Shah}, H. and {Abbey}, A.~F. and {Beardmore}, A.~P. and {Kotak}, S. and {Kamble}, N. and {Vishwakarama}, S. and {Pathare}, D.~P. and {Risbud}, V.~M. and {Koyande}, J.~P. and {Stevenson}, T. and {Bicknell}, C. and {Crawford}, T. and {Hansford}, G. and {Peters}, G. and {Sykes}, J. and {Agarwal}, P. and {Sebastian}, M. and {Rajarajan}, A. and {Nagesh}, G. and {Narendra}, S. and {Ramesh}, M. and {Rai}, R. and {Navalgund}, K.~H. and {Sarma}, K.~S. and {Pandiyan}, R. and {Subbarao}, K. and {Gupta}, T. and {Thakkar}, N. and {Singh}, A.~K. and {Bajpai}, A.},
        title = "{Soft X-ray Focusing Telescope Aboard AstroSat: Design, Characteristics and Performance}",
      journal = {Journal of Astrophysics and Astronomy},
         year = 2017,
        month = jun,
       volume = {38},
       number = {2},
          eid = {29},
        pages = {29},
          doi = {10.1007/s12036-017-9448-7},
       adsurl = {https://ui.adsabs.harvard.edu/abs/2017JApA...38...29S}
}

@ARTICLE{Kaastra...2016,
       author = {{Kaastra}, J.~S. and {Bleeker}, J.~A.~M.},
        title = "{Optimal binning of X-ray spectra and response matrix design}",
      journal = {\aap},
         year = 2016,
        month = mar,
       volume = {587},
          eid = {A151},
        pages = {A151},
          doi = {10.1051/0004-6361/201527395},
archivePrefix = {arXiv},
       eprint = {1601.05309},
 primaryClass = {astro-ph.IM},
       adsurl = {https://ui.adsabs.harvard.edu/abs/2016A&A...587A.151K}
}

@ARTICLE{Antia...2021,
       author = {{Antia}, H.~M. and {Agrawal}, P.~C. and {Dedhia}, Dhiraj and {Katoch}, Tilak and {Manchanda}, R.~K. and {Misra}, Ranjeev and {Mukerjee}, Kallol and {Pahari}, Mayukh and {Roy}, Jayashree and {Shah}, P. and {Yadav}, J.~S.},
        title = "{Large Area X-ray Proportional Counter (LAXPC) in orbit performance: Calibration, background, analysis software}",
      journal = {Journal of Astrophysics and Astronomy},
         year = 2021,
        month = oct,
       volume = {42},
       number = {2},
          eid = {32},
        pages = {32},
          doi = {10.1007/s12036-021-09712-8},
archivePrefix = {arXiv},
       eprint = {2101.07514},
 primaryClass = {astro-ph.IM},
       adsurl = {https://ui.adsabs.harvard.edu/abs/2021JApA...42...32A}
}
\bibliographystyle{aasjournalv7}



\end{document}